\documentclass[11pt]{article}

\usepackage{graphicx}
\usepackage{dcolumn}
\usepackage{bm}

\usepackage[utf8]{inputenc}
\usepackage[T1]{fontenc}
\usepackage{mathptmx}

\begin{document}


\title
{Microwave Heating and Collapse of Methane Hydrate by Molecular Dynamics Simulations}

\author{Motohiko Tanaka, Motoyasu Sato, and Shin Nakatani \\ 
Graduate School of Engineering, Chubu University, Kasugai  487-8501, Japan.}


\date{\today}
      
\maketitle      
            
\begin{abstract}
Microwave heating of methane hydrate is investigated with electrostatic molecular dynamics simulations by the SPC/E water model. The structure I of methane hydrate is constructed. When the methane hydrate with a density of 0.91 $\rm{g/cm}^3$ and temperature of 273 K is exposed to microwave electric fields, it suddenly collapses to be liquid after a certain period of irradiation. 
However, a hydrate with a five percent higher density of 0.95 $\rm{g/cm}^3$ and the same initial temperature shows no collapse as a crystal caused by the microwave. A hydrate with CO$_{2}$ guest molecules has increased kinetic energy but rapidly collapses due to the Lennard-Jones potentials. 
The methane hydrate with variable density and temperature is presented and exhibits slow heating as a crystal and an unchanging volume.

\vspace{0.2cm}
\noindent
{\small {\it 
Subjects:	Chemical Physics (physics.chem-ph) \\
arXiv:1909.01024v3, \ http://physique.isc.chubu.ac.jp}
}
\end{abstract}


%

\noindent
\section{\label{sec1}Introduction}

The natural gas resources of methane hydrate that are found in permafrost and the sea floor of our earth are drawing a lot of attention \cite{Sloan}. 
Although the production of carbon dioxide from methane hydrate is about half of that from the burning of coal and petroleum, it is still a significant contribution to greenhouse effects. Methane hydrate is a solid or liquid material and is a light electrolyte like ice. It has a density of 0.91 $\rm{g/cm}^3$ at an atmospheric pressure and 0.95 $\rm{g/cm}^3$ for an elevated pressure of 50 atm. Methane hydrate dissociates to about 220 ml of methane gas against 1 g water at 1 atm and 273 K. Methane hydrate is stable at pressures higher than 0.1 MPa at 193 K and 2.3 MPa at 273 K. 

There are three states of methane hydrates \cite{Sloan, matdata}. System I has 46 H$_2$O molecules that form $5^12$ and $5^{12} 6^{2}$ cages containing the guest molecules CH$_4$ and CO$_2$. System II has 136 H$_{2}$O molecules that form 5$^{12}$ and 5$^{12} 6^{4}$ tetragonal cages containing oxygen and other molecules. Both systems have cubic lattices. The hexagonal system H has 5$^{12}$, 5$^{12} 6^{16}$, and $4^{3} 5^{6} 6^{3}$ polyhedron cages of the C$_{6}$H$_{14}$ molecules, which exist as a hexagonal lattice.

High-pressure experiments on methane hydrate have been performed using a diamond-anvil cell \cite{Hirai}. Experiments below the melting point of ice surveyed energetically for hydrates. The stability of hydrates in the thermodynamic instability region of the ice Ih clathrate has been discussed \cite{stab-md}.
Many traditional equations of state have been utilized to describe thermo-physical properties and phase equilibrium \cite{Soave}.
The multi-scale phase diagram of the Gibbs-Helmholtz constrained equation of state for methane hydrate has been tabulated by density for given pressures and temperatures  \cite{lucia2010,density-tab}.

The diffusion coefficients and dielectric relaxation properties of water, i.e., the response of electric dipoles to a given initial impulse, have been studied theoretically \cite{yamag}.
The heating and diffusion of water under high-frequency microwaves and infrared electromagnetic waves have been investigated by molecular dynamics simulations using
elaborated point-charge models \cite{engl}.
Molecular dynamics simulation of the ice nucleation and growth process leading to water freezing has been executed \cite{matsu}.

Molecular dynamics simulations have been conducted concerning the microwave heating of water, ice, and saline solution \cite{cite01}. 
They have shown that: (i) water in the liquid phase is heated via the rotational excitation of water electric dipoles, which is delayed from the microwave electric field, and absorbed the total microwave power; (ii) hot water gains significantly less heat than the water at room temperature because of smaller phase lags due to less friction; (iii) water in the ice phase is scarcely heated because the electric dipoles can not rotate due to the tightly hydrogen-bonded ice crystal; (iv) dilute saline solution gains significantly more heat than pure water because of the rapid heating of salt ions, especially that of the large salt ions Cl$^{-}$ and Na$^{+}$.

Molecular dynamics employing the density functional method (DFT) to simulate the THz range of electromagnetic wave have been constructed \cite{cite02}.
They have shown, by the self-consistent atomic forces \cite{Siesta}, that: (i) liquid water molecules in the electric field has excited rotational motions, as water molecules in the cages can not make free translation motions; (ii) the electron energy is about twice the kinetic energy of the water molecules, which results from the forced excitation of the molecules by the electromagnetic THz external field.

Regarding the methane hydrates as natural resources, there are several methods for collection from the sea floor. A very important question arises concerning heating of the methane hydrates. One might ask whether heating with microwaves could be continued beyond 273 K as a crystal or liquid ? 

The rest of this paper is organized as follows. The methodology of electrostatic molecular dynamics with boundary conditions is given in Section \ref{sec2}. Modeling of present simulations is given in Section \ref{sec3.1}, and heating and collapse of methane hydrate are shown in Section \ref{sec3.2}. A simulation with constant pressure is shown in Section \ref{sec3.3}. A summary and conclusion are provided in Section \ref{sec4}. The equations of long-range Coulombic interactions are explained in Appendix A.

\section{\label{sec2}Molecular Dynamics Methodology by SPC/E Model}

Crystal structures are a part of the strongly coupled systems in solid-state physics \cite{Kittel}, $\Gamma = e^{2}/\epsilon ak_{B}T > 1$, where $e$ is charge,  $\epsilon$ is the electrical constant, $a$ is an inter-particle distance, and $k_{B}T$ is the thermal energy. Four basic quantities are used to derive the Newtonian equation of motion for water molecules: the time $\tau_{0}= 1 \times 10^{-14} \ \rm{s}$, the length 1 \AA = $1 \times 10^{-8} \ \rm{cm}$, mass of water $M_{0} = 3.0107 \times 10^{-23} \ \rm{g}$, and electronic charge $e = 4.8033 \times 10^{-10} \ \rm{esu}$ \ ($1.6022 \times 10^{-19} \ \rm{C}$ in the international units system). 
Then, one has the equations \cite{cite01,cite03}, 
\begin{eqnarray}
M_{i} \frac{d\vec{v}_{i}}{d \tau} &=&  -\nabla \left\{ 
\Phi_{F} (\vec{r}_{i}) + 4 \epsilon_{ij} \left[ \Bigl( \frac{\sigma_{ij}}{r_{ij}} \Bigr)^{12} -\Bigl( \frac{\sigma_{ij}}{r_{ij}} \Bigr)^{6} \right]
\right\}  + q_{i}E \sin \omega t \ \hat{x}, 
\label{eq1} \\
\frac{d\vec{r}_{i}}{d \tau} &=& \vec{v}_{i}.
\label{eq2}
\end{eqnarray}
that define the equations of motion.
The first term of the right-hand side of Eq.(\ref{eq1}) is the Coulombic potential $\Phi_ {F}(\vec{r}_{i})$, and the second term is the Lennard-Jones potential. Here, $\vec{r}_{i}$ and $\vec{v}_{i}$ are the position and velocity of $i$-th molecule, respectively, $M_{i}$ and $q_{i}$ are the mass and charge, respectively, $\tau$ is the time, and $\nabla$ is the space derivative. The quantity $\vec{r}_{ij}= \vec{r}_{i} -\vec{r}_{j}$ is the particle spacing between the $i$-th and $j$-th molecules, and $\epsilon_{ij}$ and $\sigma_{ij}$ are the two-particle interaction potential and size of the Lennard-Jones potential, respectively. 
The external electric field points to the $x$ direction and has the form $\sin \omega t$, and frequency $f$ of $\omega= 2 \pi f$.
The time step is $\Delta \tau= 0.1 \tau_{0}$ (i.e., $1 \times 10^{-15} \rm{s}$).
In a time marching fashion, the current step of $\vec{r}_{i}$ and $ \vec{v}_{i}$ is forwarded to the next time step. When a sufficient amount of time has elapsed, one analyzes the time development.

To represent the crystal system with high accuracy, one has to separate the Coulombic forces $\vec{F}(\vec{r}_{i})= - \nabla \Phi_{F}(\vec{r}_{i})$ that occur in the short-range and the long-range interactions \cite{Ewald, Kolafa, Frenkel},
\begin{eqnarray}
\vec{F}(\vec{r}_{i}) &=& \vec{F}_{SR}(\vec{r}_{i}) +\vec{F}_{LR}(\vec{r}_{i}).
\end{eqnarray}
The short-range interactions are written as,
\begin{eqnarray}
\vec{F}_{SR} (\vec{r}_{i}) &=& \sum_{j=1}^{N} q_{i}q_{j} \Bigl[ \Bigl( \frac{\rm{erfc}(\it{r}_{ij})}{r_{ij}} +\frac{2\alpha}{\sqrt \pi} \Bigr) \exp(-(\alpha r_{ij})^2)/r_{ij}^{2} \Bigr] \vec{r}_{ij},
\label{Shortrg} 
\end{eqnarray}
where the Gauss complimentary error function is
\begin{eqnarray}
\rm{erfc}(\it{r})= \frac{2}{\sqrt \pi} \int_{r}^{\infty} \exp(-t^{2}) dt. 
\label{contgaus}
\end{eqnarray}
The $\alpha$ value, a minimization factor, is discussed later.

A primary factor in the long-range interactions is the charge density, $\rho(\vec{r}_{i})= \sum_{j} q_{j}S(\vec{r}_{i}-\vec{r}_{j})$, which is the near-site grid sum with $S(\vec{0})= 1$, $S(\infty)\rightarrow 0$. Then, the grid summation is converted to the k-space by a Fourier transform $FT^{-1}[...]$.
Here, $\rho(\vec{r}) \rightarrow \rho_{k}(\vec{k})$ with $\vec{k}= 2 \pi \vec{n}/ \it{L}$, $n$ the integers $\ge 0$, and $L$ the length. The inverse Fourier transform to return to the coordinate space is executed by the folding operations $FT[...]$,
\begin{eqnarray}
 \vec{F}_{LR}(\vec{r}_{i}) &=& -FT \Bigl[ i \ q_{i} 
   (dn(n_x),dn(n_y),dn(n_z)) G(n_x,n_y,n_z) \ \rho_{k}(\vec{k}) \Bigr] , \\
  && dn(n_{\gamma})= n_{\gamma} - dnint(n_{\gamma}/M_{\gamma}) M_{\gamma}
 \hspace*{0.3cm} (\gamma= x,y,z).
\label{Longrg} 
\end{eqnarray}
The expressions for the $G(n_x,n_y,n_z)$, $\vec{K}(n_x,n_y,n_z)$ and $\Delta(n_x,n_y,n_z)$ functions are given in Appendix A. The $\alpha$ value is determined by minimizing the errors of both the short-range and long-range interactions of the electric fields \cite{Deserno}. The value is $\alpha=0.245$ for the total number of $3^3$ methane hydrates. 

The constrained dynamics procedure called "Shake and Rattle algorithm" is used to maintain the bond lengths and angles \cite{SandR}. The pre- and post-iterations of Coulombic forces are required in the simulation code. An accuracy of at least five digits for each molecule is achieved, and the time advancement is made to the next step.

It is very important that the Symplectic Integrator (SI) scheme \cite{kang, forest, okabe} is applied in the differential equations of Eq.(\ref{eq1})-Eq.(\ref{Longrg}).
The one-sided energy drift in the Nordsieck-Gear scheme \cite{gear} does not exist in the differential equations of the SI scheme. The simulation is controlled inversely to the order of the scheme. The second order scheme is prescribed with one Coulombic force calculation in each time step.


\begin{table}[h]
\caption{The series of simulations, fixed density ($\rm{g/cm}^3$), microwave heating rate ($W_{0}/\tau_{0}$), evaluation, and guest molecules for the constant volume case.}
\label{Table-1}
  \centering 
  \vspace{-0.2cm}
  \begin{tabular}{cllll} 
\hspace{0.3cm} \\ \hline
 \hspace{0.2cm} series & \hspace{0.2cm} density & \hspace{0.2cm} heating rate & \hspace{0.2cm} evaluation & \hspace{0.2cm} guest molecules \hspace{0.1cm} \\ \hline
 A1 & \hspace{0.2cm} $0.91 \rm{g/cm}^3$ & \hspace{0.2cm} $4.2 \times 10^{-7} W_{0}/\tau_{0}$ & \hspace{0.2cm} $1.05 \times 10^{6} \tau_{0}$ \  collapsed &  \hspace{0.2cm} $\rm{CH}_4$ \\ 
 A2 & \hspace{0.2cm} $0.95 \rm{g/cm}^3$ & \hspace{0.2cm} $1.7 \times 10^{-7} W_{0}/\tau_{0}$ & \hspace{0.2cm} $1.5 \times 10^{6} \tau_{0}$ \  remains crystal & \hspace{0.2cm} $\rm{CH}_4$ \\ 
 A3 & \hspace{0.2cm} $0.91 \rm{g/cm}^3$ & \hspace{0.2cm} $3.9 \times 10^{-7} W_{0}/\tau_{0}$ & \hspace{0.2cm} $6.6 \times 10^{5} \tau_{0}$ \ collapsed &\hspace{0.2cm} $\rm{CO}_2 $, $\rm{CH}_4 $  
\hspace{0.3cm} \\ \hline

  \end{tabular}
\end{table}

\begin{table}[h]
\caption{The series of simulation, initial density ($\rm{g/cm}^3$), microwave heating rate ($W_{0}/\tau_{0}$), evaluation, and guest molecules for the variable density case.}
\label{Table-2}
  \centering 
  \vspace{-0.2cm}
  \begin{tabular}{cllll} 
\hspace{0.3cm} \\ \hline
 \hspace{0.2cm} series & \hspace{0.2cm} initial density & \hspace{0.2cm} heating rate & \hspace{0.2cm} evaluation & \hspace{0.2cm} guest molecules \hspace{0.1cm} \\ \hline
 B1 & \hspace{0.2cm} $0.95 \rm{g/cm}^3$ & \hspace{0.2cm} $1.3 \times 10^{-7} W_{0}/\tau_{0}$ & \hspace{0.2cm} $5.7 \times 10^{5} \tau_{0}$ \ remains crystal &  \hspace{0.2cm} $\rm{CH}_4$  
 \hspace{0.3cm} \\ \hline

  \end{tabular}
\end{table}

\section{\label{sec3}Simulations of Methane Hydrate by SPC/E Model}
\subsection{\label{sec3.1}Modeling}

To model water molecules, there is the three-body SPC/E model \cite{Berendsen} and the four-body TIP4P model \cite{Jorgensen}. The SPC/E model is chosen because of its simple and effective representation of electrostatic effects. The atomic distance of the oxygen-hydrogen molecule is 1.00 \AA, the respective charges of oxygen and hydrogen are $q_{O}=-0.848 \ e$ and $q_{H}=0.424 \ e$, and the angle of H-O-H is $\theta= 109.47^{\circ}$.
The Lennard-Jones parameters are made for oxygen only, which are $\sigma_{O}=3.17$ \AA \ and $\epsilon_{O}= 0.65 \ \rm{kJ/mol}$. 

The initial structure of methane hydrate is installed by the Genice program on the Linux system \cite{cite04}. The size of structure I methane hydrate has about 12  \AA \ as the crystal structure.  
A total of 3$^3$ methane hydrates exists in the three-dimensional system. A system of guest molecules of $\rm{CH}_4$ or $\rm{CO}_2$, which are the united atoms, is used with different densities of the guest molecules. The size and Lennard-Jones coefficient for the CH$_4$ molecule are $\sigma_{CH_4}=3.82$ \AA\ and $\epsilon_{CH_4}=0.39 \ \rm{kJ/mol}$, and those for carbon dioxide are $\sigma_{CO_2}=4.00$ \AA \ and $\epsilon_{CO_2}=0.53 \ \rm{kJ/mol}$, respectively \cite{AV}.

In order to perform the molecular dynamics simulation, the frequency is set to 10 GHz ($f=1 \times 10^{4} /\tau_{0}$) and the electric field is set to 0.3 V/\AA. \  
The rationale behind setting these parameters will be discussed in the final section.  

For numerical execution of the molecular dynamics simulations, a Fujitsu FX100 Supercomputer (52 ranks$\times$16 processors, 8 thread) is utilized. The computations took $1.57 \times 10^{-2}$ s per time step, and thus the run of $5.5 \times 10^6$ time steps requires 24 hours of computation.

\begin{figure}
\centering
  \includegraphics[width=4.5cm]{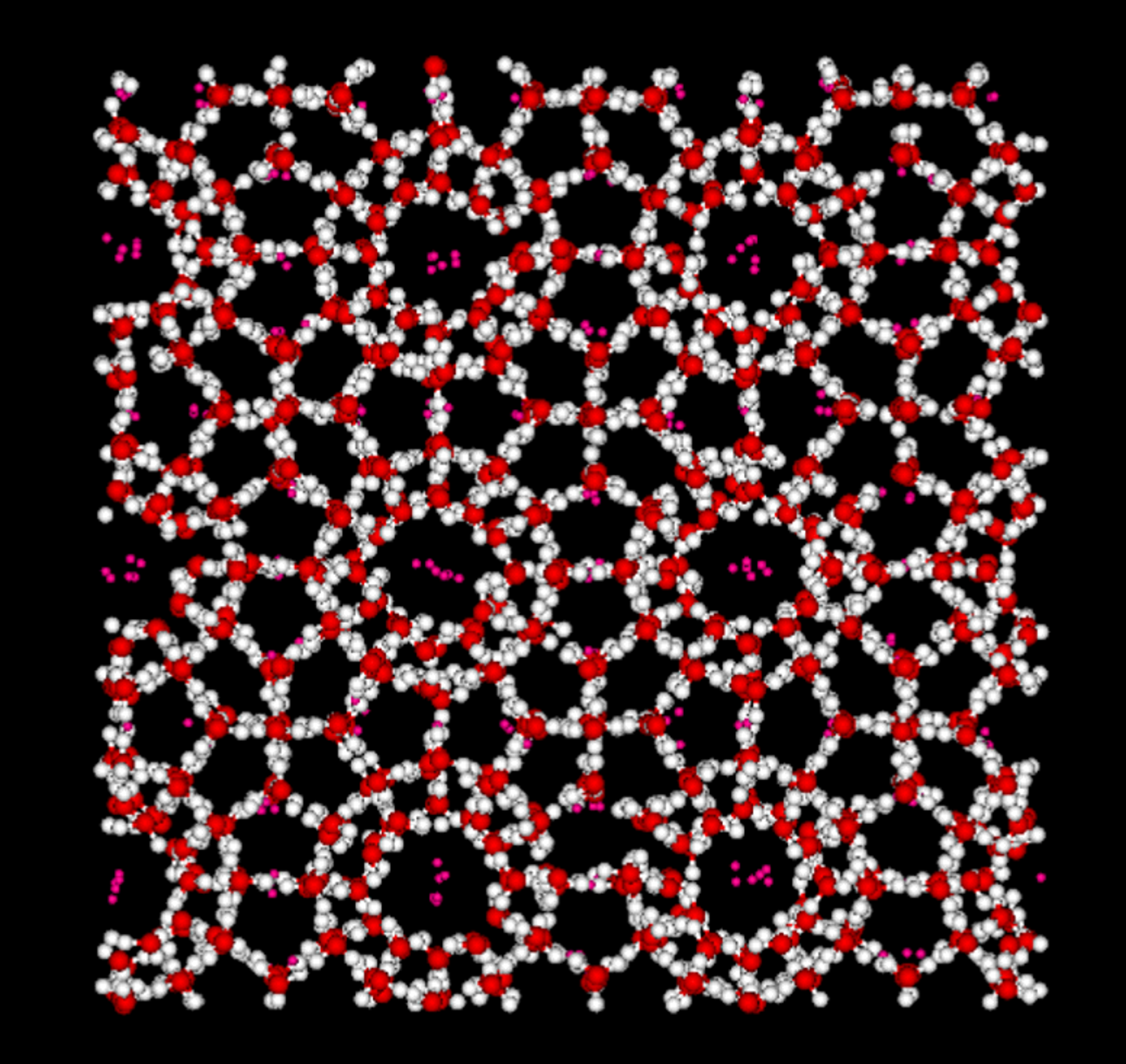} 
\hspace*{0.2cm}
  \includegraphics[width=4.38cm]{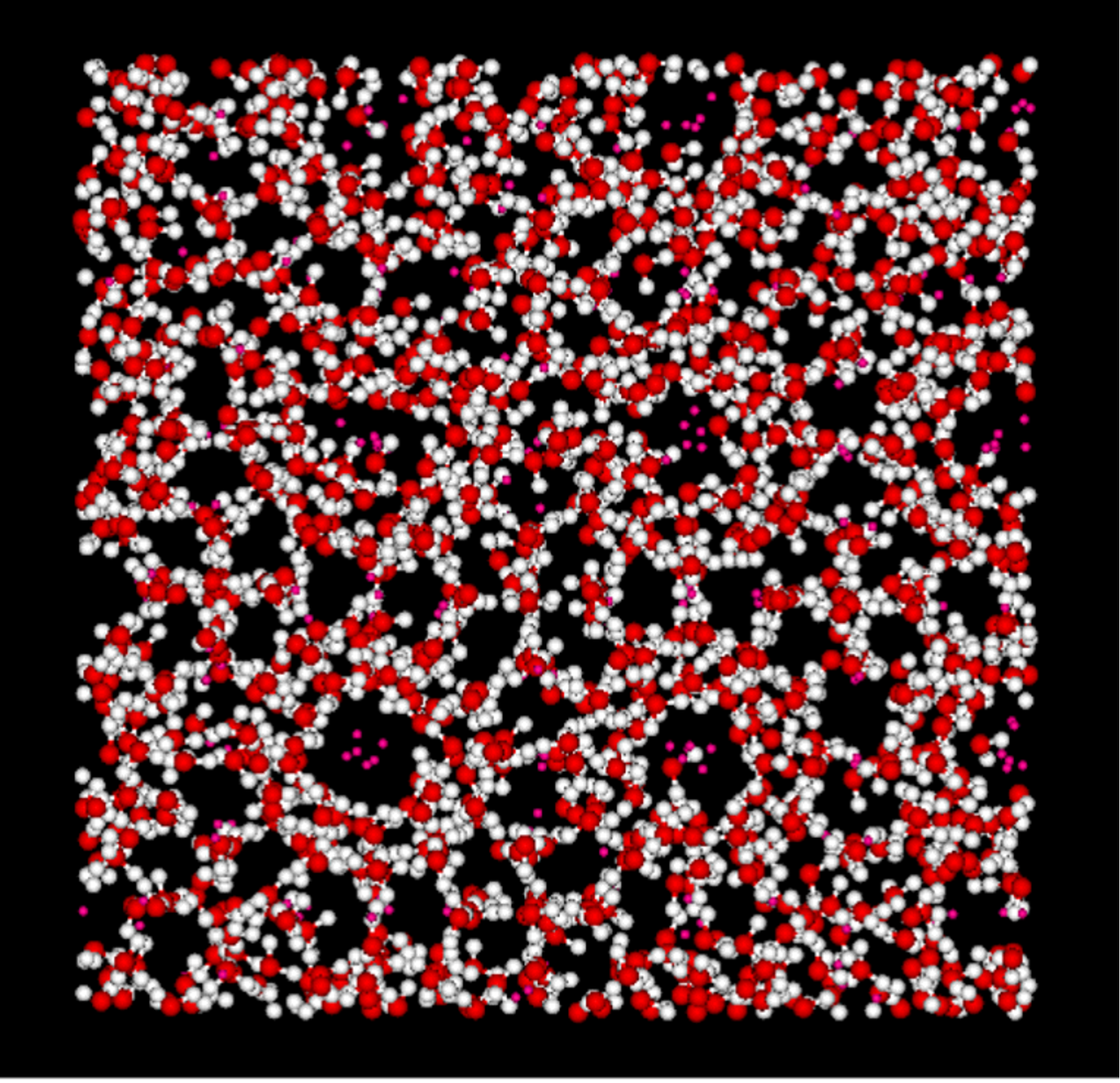} 

\caption{
(a) The initial crystal of the H (white), O (red), and CH$_4$ (small red) molecules having a density of 0.91 g/cm$^3$ and an initial temperature of 273K, 
(b) distortion of the methane hydrate to be liquid under an applied microwave field observed at the time $\tau = 1.1 \times 10^{6} \tau_{0}$.
}
\label{mhr023plot}
\end{figure}

\begin{figure}
\centering
  \includegraphics[width=8.0cm,clip]{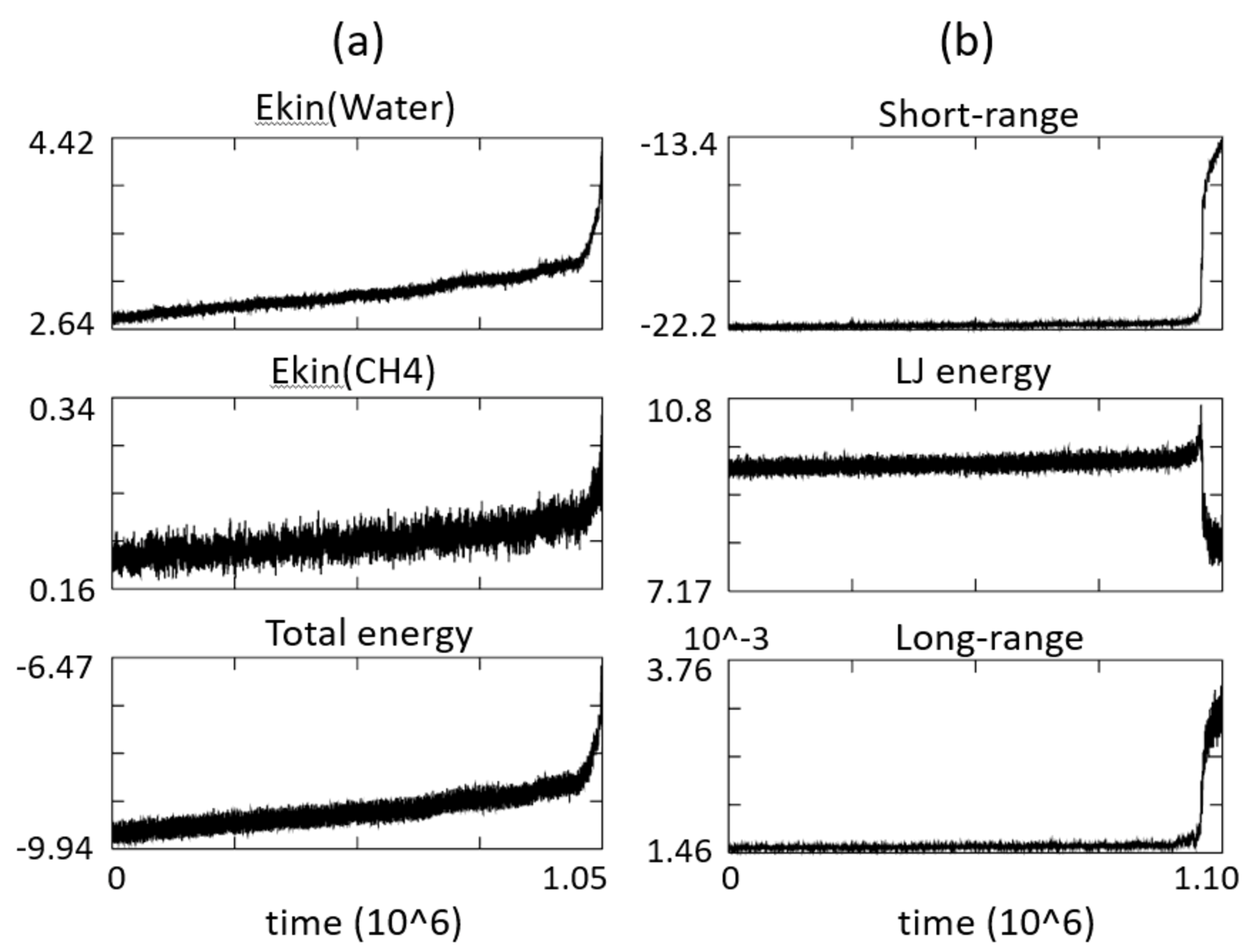} 

\caption{
Microwaves are applied to a methane hydrate with a density of 0.91 g/cm$^3$ and a temperature of 273 for $\tau > 0$.
(a) The kinetic energy of water, that of CH$_4$, and the total energy of the system by $\tau= 1.05 \times 10^{6} \tau_{0}$ (left, top to bottom, respectively). 
(b) The longer timescale plot of the short-range interaction energy, the Lennard-Jones potential energy, and long-range interaction energy to $\tau = 1.10 \times 10^{6} \tau_{0}$ (right, top to bottom, respectively).
The abscissa is linearly scaled in all plots hereafter. 
The kinetic energy of water increases with a time rate of $\Delta W/\Delta \tau= 4.2 \times 10^{-7} W_{0}/\tau_{0}$, and the short and long-range energies eventually collapse to be liquid at the time $\tau \cong 1.05 \times 10^{6} \tau_{0}$, whereas the Lennard-Jones energy decreases at the same time. 
}
\label{mhr023kin}
\end{figure}

\begin{figure}
\centering
  \includegraphics[width=7.0cm]{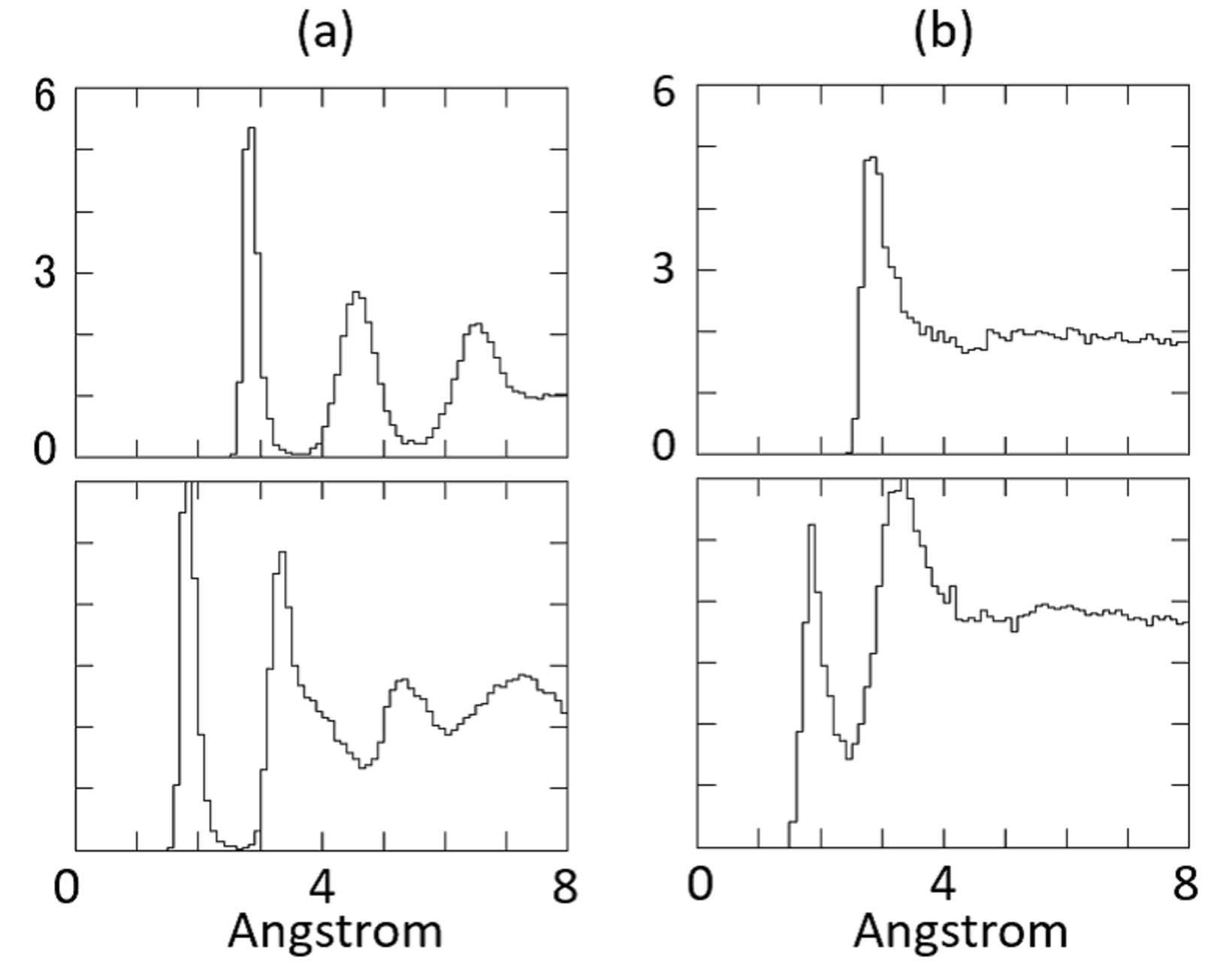} 

\caption{
The pair distribution functions between the O-O atoms (top) and O-H atoms (bottom) of the methane hydrate for the density 0.91 g/cm$^3$. The time is (a) before the collapse at $\tau=1.05 \times 10^{6} \tau_{0}$, and (b) after the collapse at $\tau =1.06 \times 10^{6} \tau_{0}$. 
The abscissa of the distribution functions is (a) $1.39 \times 10^{-2}$ (left), and (b) $8.75 \times 10^{-3}$ (right).
The three peaks for the O-O atoms in (a) show that the atoms have almost aggregated by the collapse in (b), and the two giant peaks in (b) indicate the O-H atoms.
}
\label{Fig_FV2}
\end{figure}

\subsection{\label{sec3.2}Heating and Collapse of Methane Hydrate}

The heating of methane hydrates is investigated under the application of microwave fields. In the heating described in Section \ref{sec3.2}, the volume is assumed to be constant. Three species of hydrates are used in Table \ref{Table-1}: (i) a normal pressure case (1 atm, Run A1), (ii) a high pressure methane hydrate at 50 atm (Run A2), and (iii) carbon dioxide hydrate at 1 atm (Run A3). 
The density is 0.91 $\rm{g/cm}^{3}$ for the first case \cite{Sloan, density-tab}, 0.95 $\rm{g/cm}^{3}$ for the second case \cite{matdata}, and 0.91 $\rm{g/cm}^{3}$ for the CO$_2$ hydrate in the third case. Each run has an initial temperature of 273 K of the crystal. 

The kinetic energy of the $s$-th molecule is $W_{s}= (1/2)M_{s} \vec{v}_{s}^{2} +(1/2)I \omega^{2}$, where $\vec{v}_{s}$ and $I$, respectively, are its velocity and moment of inertia.
The translational and rotational motions are included for the water molecules. The mass of water $M_{\rm{H_2 O}}$ is set to unity, $\vec{v}_{s0}$ is the initial velocity, and  $\vec{v}_{w0}$ is set for the initial water molecules. A dry run that does not include any microwaves is first executed. It shows a very small and non-increasing drift, as was described at the end of Section 2. This basis is then assumed for all other runs.

Microwaves are applied in Run A1. The initial state of the hydrogen, oxygen, and CH$_{4}$ (unified atom) molecules is shown in Fig.\ref{mhr023plot}(a). With the microwave electric field on, the kinetic energies of the water and CH$_4$ molecules increase over time, as shown in Fig.\ref{mhr023kin}(a) and (b). The total energy of the system, i.e. the kinetic energy and the Coulombic interaction energy, is depicted in Fig.\ref{mhr023kin}(c).
The increase in the energy of the water is $\Delta W/\Delta \tau \cong 4.2 \times 10^{-7} W_{0}/\tau_{0}$.
The CH$_{4}$ molecules, without any charges, are inert to microwaves, but they interact with water molecules. The energy increases in the Lennard-Jones potential, which was close to the water case. 

The nonlinear growth of methane hydrate occurs for the time $\tau \ge 7 \times 10^{5} \tau_{0}$. The explosive growth suddenly collapses and methane hydrate turns to be the liquid phase at $\tau \cong 1.05 \times 10^{6} \tau_{0}$. 
The collapse is depicted in the particle plot of Fig.\ref{mhr023plot}(b). The water molecules scatter as early as time $\Delta \tau \cong 500 \tau_{0}$.
The explosive phase is depicted in Figs.\ref{mhr023kin}(a)-(c) at $\tau= 1.05 \times 10^{6} \tau_{0}$, and the collapse in Fig.\ref{mhr023kin}(d)-(f) at $\tau= 1.10 \times 10^{6} \tau_{0}$. The Lennard-Jones energy decreases chaotically in Fig.\ref{mhr023kin}(e). The temperature increase before the microwave collapse for Run A1 is $\Delta T \cong 35 $ deg.

The pair distribution functions of different O-O and O-H atoms are shown in Fig. \ref{Fig_FV2} for the methane hydrate with a density of $0.91 \rm{g/cm}^3$.
They show the distribution functions that exist before and after the collapse occurring around $\tau \cong 1.05 \times 10^{6} \tau_{0}$.
Well separated peaks in the 8 \AA \ regions can be identified as a crystal before the collapse on the top and bottom of the left side column. However, one has only one peak after the collapse as liquid, which is entirely buried in the $r >$ 3 \AA \ region of the O-O distribution function. Two peaks are seen with curtains in the O-H functions of the right column.

\begin{figure}
\centering
  \includegraphics[width=4.5cm,clip]{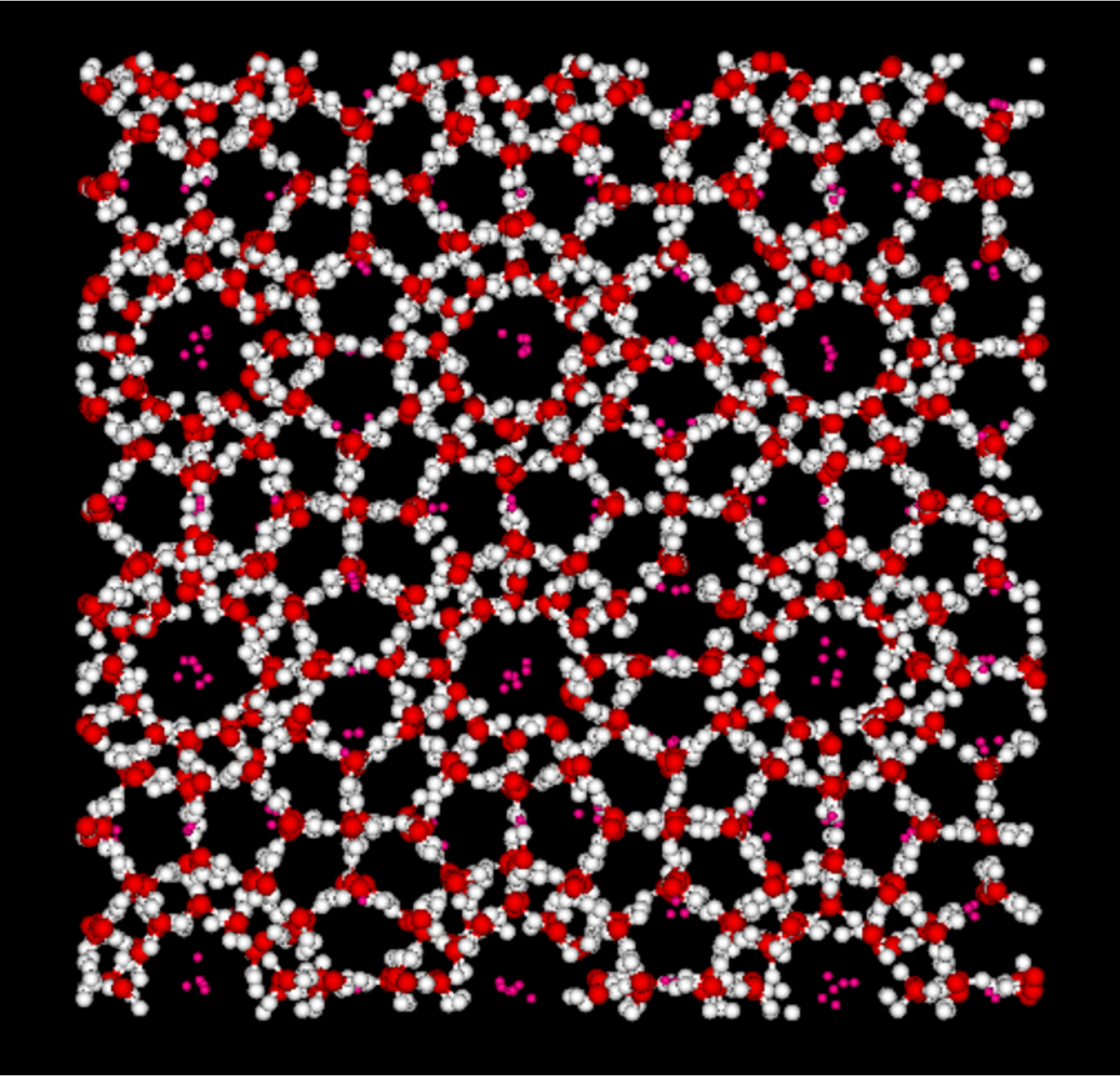} 

\caption{
The density 0.95 g/cm$^3$ and the initial temperature 273 K lead to the continuous heating of methane hydrate as a crystal, at $\tau = 1.5 \times10^{6} \tau_{0}$. The legend is H (white), O (red), and CH$_4$ (small red). 
}
\label{mhr025plot}
\end{figure}

\begin{figure}
\centering
  \includegraphics[width=8.0cm,clip]{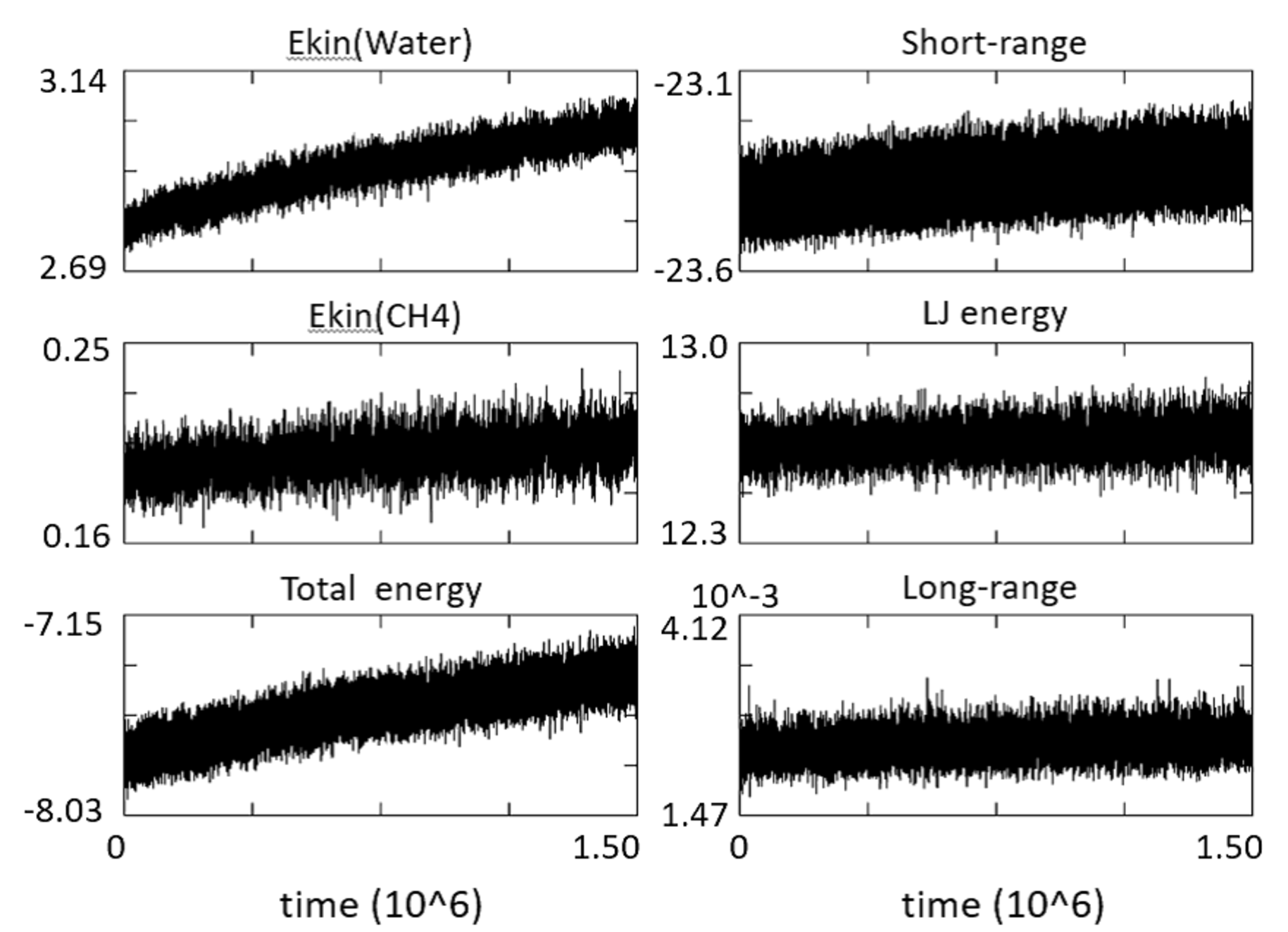} 

\caption{
The kinetic energy of water, that of CH$_4$, and the total energy of the system (left, to and bottom), the short-range interaction energy, the Lennard-Jones energy, and long-range interaction energy (right, top and bottom) for microwave application of the density 0.95 g/cm$^3$. The abscissa is linearly scaled. 
The kinetic energy is increased in a continuous manner as a crystal at $\tau = 1.5 \times10^{6} \tau_{0}$, and the heating rate is $\Delta W/\Delta \tau= 1.7 \times 10^{-7} W_{0}/\tau_{0}$. 
}
\label{mhr025kin}
\end{figure}

Alternatively, a five percent higher density of 0.95 $\rm{g/cm}^{3}$ and an initial temperature of 273 K are simulated in Run A2. This corresponds to 0-5 $^{\circ}$C and 50 atmospheres \cite{matdata}.
A time of $\tau \cong 1.5 \times 10^{6} \tau_{0}$ is executed with the same electric field of the microwave used in Run A1, and the particle plot is shown in Fig.\ref{mhr025plot}.
The kinetic energy of water in Fig.\ref{mhr025kin}(a) increases at a very slow rate of $\Delta W/\Delta \tau= 1.7 \times 10^{-7} W_{0}/\tau_{0}$. 
A collapse similar that seen for the methane hydrate in Run A1 does not occur but continues as a crystal, as shown in the figures.

\begin{figure}
\centering
  \includegraphics[width=4.5cm,clip]{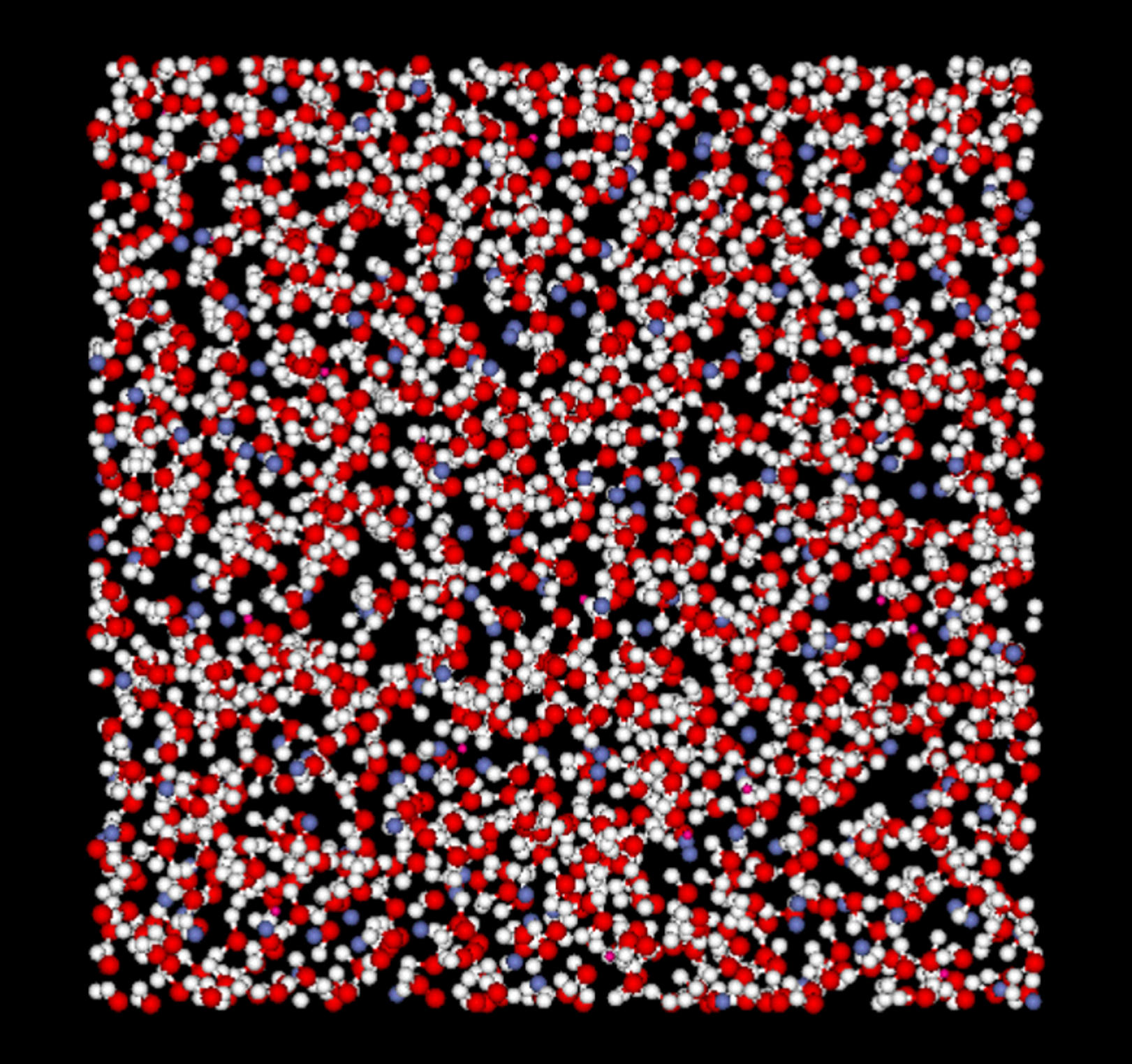}
\caption{
The density 0.91 g/cm$^3$ and the initial temperature 273 K become collapsed to be liquid by the carbon dioxide molecules at $\tau \cong 6.6 \times 10^{5} \tau_{0}$. The legend is H (white), O (red), and CO$_2$ (small blue). }
\label{mhrcm3plot}
\end{figure}

\begin{figure}
\centering
  \includegraphics[width=8.0cm,clip]{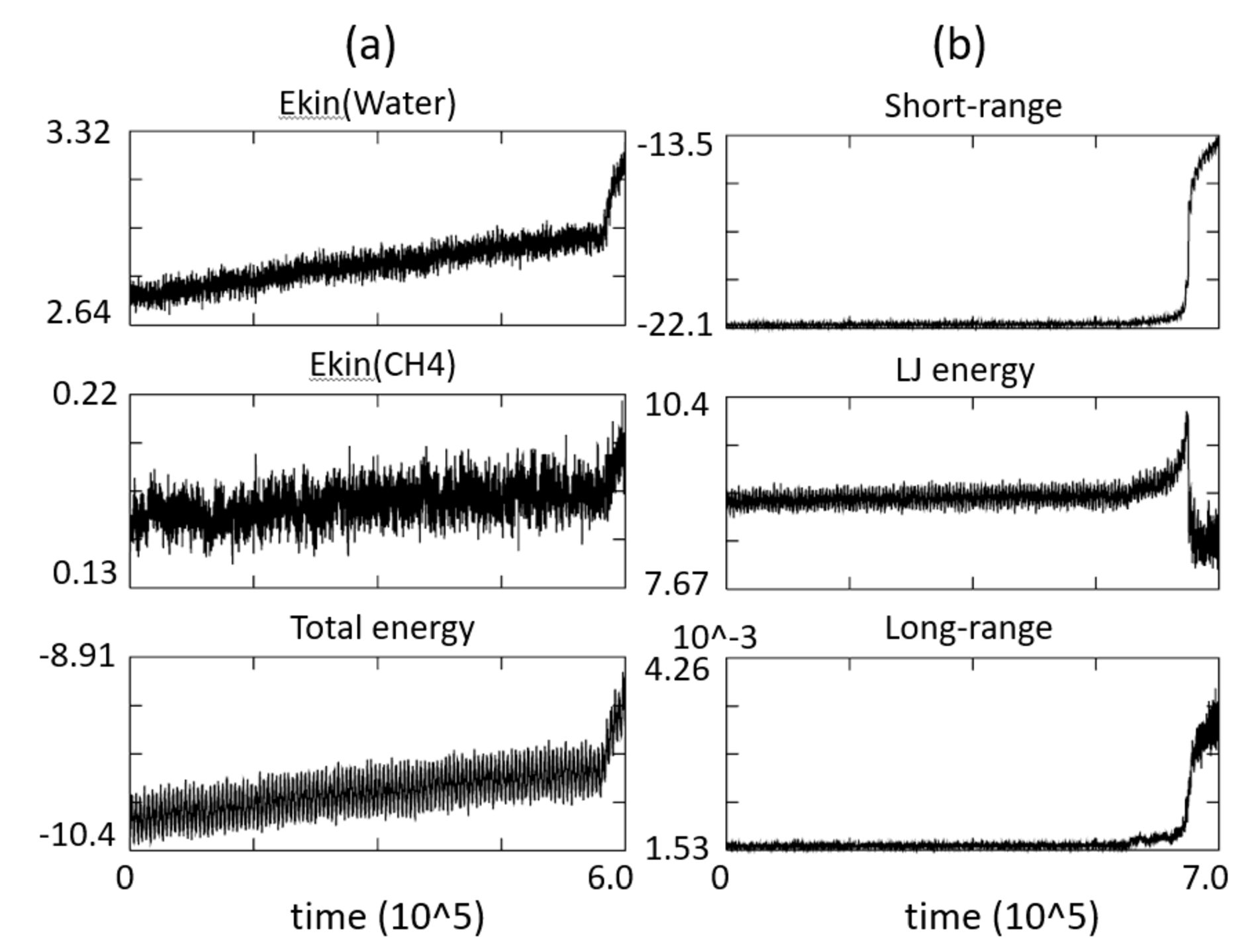} 
\vspace{-0.3cm}
\caption{
Simulation of 195 carbon dioxide and 22 methane molecules mixed with a density of 0.91 g/cm$^3$ and run at an initial temperature of 273 K. (a) The kinetic energy of water, that of CH$_4$, and the total energy of the system for $\tau= 6 \times 10^{5} \tau_{0}$ (left, top to bottom, respectively).
(b) The longer timescale plot of the short-range interaction energy, the Lennard-Jones potential energy, and long-range interaction energy for $\tau = 7 \times 10^{5} \tau_{0}$ (right, top to bottom, respectively).
The abscissa is linearly scaled. The kinetic energy of water increases at the rate $\Delta W/\Delta \tau= 3.9 \times 10^{-7} W_{0}/\tau_{0}$, but collapses to be liquid at $\tau \cong 6.6 \times 10^{5} \tau_{0}$. 
}
\label{mhrcm3kin}
\end{figure}

\begin{figure}
\centering
  \includegraphics[width=8.0cm,clip]{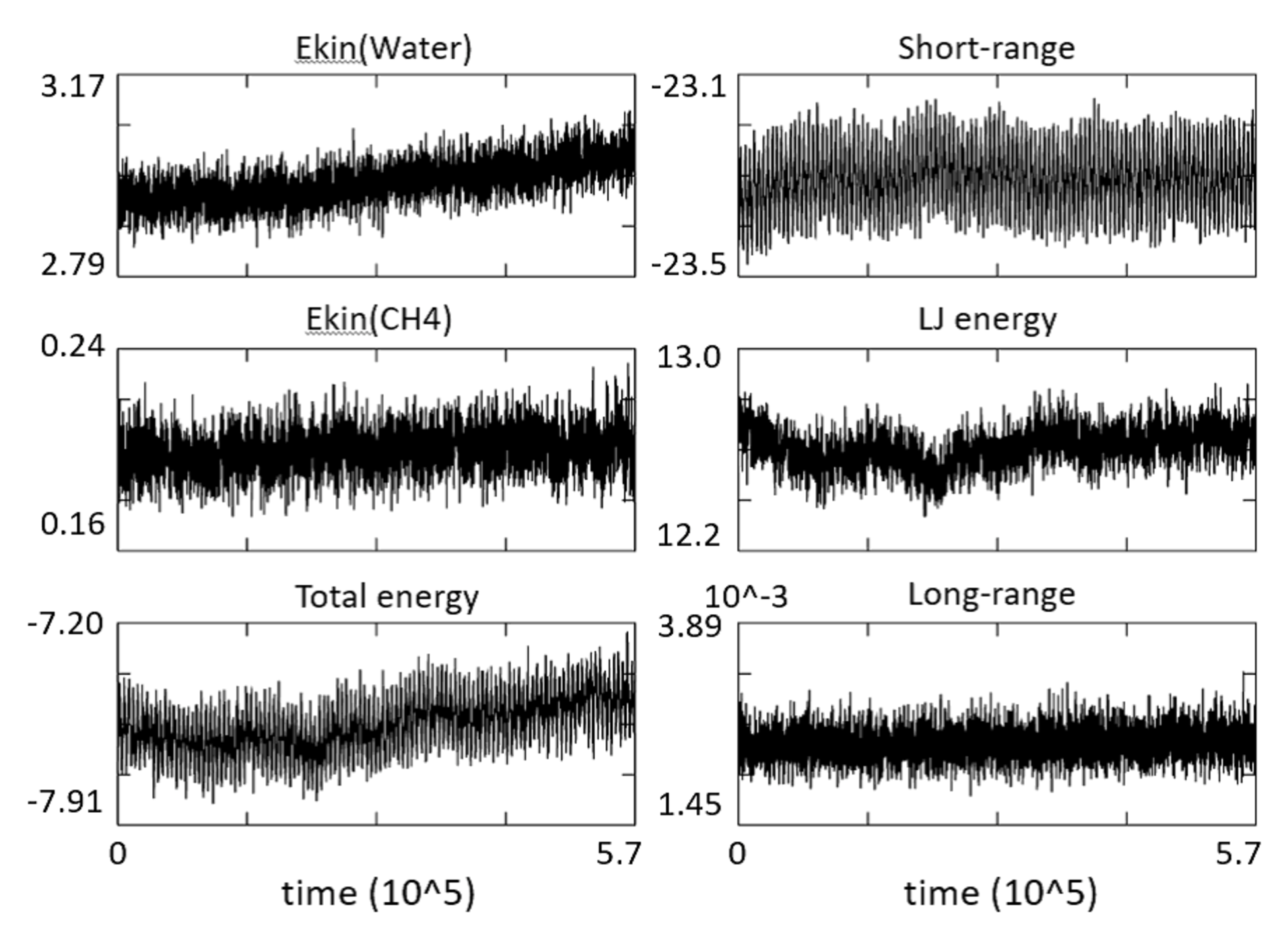} 
%
\caption{
The change of the initial density of 0.95 g/cm$^3$ and temperature of 273 K. (a) The kinetic energy of water, that of CH$_4$, and the total energy of the system (left, top to bottom, respectively). (b) The detailed plots of the short-range interaction energy, the Lennard-Jones potential energy, and the long-range interaction energy for $\tau = 5.7 \times 10^{5} \tau_{0}$ (right, top to bottom, respectively). 
The kinetic energy of water is $\Delta W/\Delta \tau= 1.3 \times 10^{-7} W_{0}/\tau_{0}$ as a crystal, which is smaller than the constant volume case of Run A2.
}
\label{mhr036kin}
\end{figure}

A simulation of $3^3$ methane hydrate is executed where 194 carbon dioxide and 22 methane molecules are mixed in Fig. \ref{mhrcm3plot} (Run A3). The weight of the guest molecule CO$_{2}$ (i.e., 44) is about three times heavier than that of the methane molecule CH$_{4}$. 
%
The kinetic energies of water and CO$_2$ molecules also increase in Fig. \ref{mhrcm3kin}.
The present hydrate is heated at a slightly slower rate of $\Delta W/\Delta \tau \cong 3.9 \times 10^{-7} W_{0}/\tau_{0}$, compared to the methane hydrate of Run A1 with a density of 0.91 g/cm$^{3}$.
It collapses at the time $\tau \cong 6.6 \times 10^{5} \tau_{0}$.
The heating of the hydrate with CO$_{2}$ molecules is $\Delta W= 2.8 \times 10^{-1} W_{0}$, which is about a half that of the methane hydrate as $\Delta W= 4.5 \times 10^{-1} W_{0}$ of Run A1. 
%
The CO$_2$ molecules are by 0.53$\rm{kJ/mol}$ more active at the Lennard-Jones potential than the CH$_4$ molecules of 0.39$\rm{kJ/mol}$.
This may contribute to a shorter life of carbon dioxide hydrate.

\subsection{\label{sec3.3}Heating due to Density and Temperature Changes}

In this subsection, the methane hydrate with the initial density of 0.95 $\rm{g/cm}^3$ and temperature of 273 K is advanced to a new density and temperature under the microwave in Table \ref{Table-2}. When a deviation of the kinetic energy of water exceeds $\pm 0.1 \times 10^{-4} W_{0}$ in the interval of $\Delta \tau= 2000 \tau_{0}$, the system length $L$ is changed by a $\pm 0.005$ \AA \ stride (double sign correspondence). If the deviation is within the threshold, the system size is not altered for that interval. 

Figure \ref{mhr036kin} shows the kinetic energy of water, that of CH$_4$, and the total energy of the system at $\tau=  5.7 \times 10^{5} \tau_{0}$ in the left column. The details of the short-range, Lennard-Jones and long-range interaction energies are shown in the right column. The Coulombic energy gradually increases and decreases, while the Lennard-Jones energy goes in opposite directions. The sum of the total energy, i.e. the kinetic energy and the total interaction energy, thus follows the same trend in Fig.\ref{mhr036kin}(c).
The system size stays within the $35.60 \pm 0.05$ \AA, and continues as a crystal. 
The kinetic energy of water increases with $\Delta W/\Delta \tau= 1.3 \times 10^{-7} W_{0}/\tau_{0}$. This value is about 70\% smaller than that of the volume fixed case of Run A2. This results in the slow process of the methane hydrate heating.

\section{\label{sec4}Summary}

Methane hydrate was simulated by molecular dynamics with periodic boundary conditions. The SPC/E model was used to show their results compared with the TIP5P-Ewald model \cite{MDTip5p}.
The normal density of 0.91 g/cm$^3$ represented sea level conditions, and the density of 0.95 g/cm$^{3}$ corresponded to the high-pressure conditions of 50 atm and 273 K. 
The microwave electric field was applied while the volume was held constant. The methane hydrate collapsed to be liquid for the normal density, and the temperature increase before the collapse was $\Delta T \cong 35 $ deg. 
Whereas it continued to be heated for the higher density as a crystal. 
The simulation of carbon dioxide molecules starting at 273 K was executed but was collapsed to be liquid shortly compared to the methane hydrate.

The methane hydrate was dynamically adjusted to density and temperature by the microwave process. It was found to be heated by the microwave, and continued as a crystal.
But, the system size stayed almost constant, which resulted in the slow process of the methane hydrate heating.


\vspace{0.7cm}
\noindent
{\large {\bf Acknowledgments}}

\vspace*{0.3cm}
The author (M.T.) is grateful to Professor M. Matsumoto of Okayama University for the initial configuration of methane hydrate. The computation is performed by Fujitsu FX100 Supercomputer of National Institute of Fusion Science, Japan.


    \renewcommand{\theequation}{A.\arabic{equation}}
    \setcounter{equation}{0}

\vspace{1.0cm}
\noindent
{\large \label{append-A}
{\bf Appendix A: Long-range Coulombic Interactions}
}

\vspace{0.2cm}
The $G(n_x,n_y,n_z)$, $\vec{K}(n_x,n_y,n_z)$ and $\Delta(n_x,n_y,n_z)$ functions for the long-range Coulombic interactions are written in the periodic boundary conditions as,  
\begin{eqnarray}
  && \hspace*{-1.6cm}
   G(n_x,n_y,n_z) = (2 M_x M_y M_z/L^{2}) \times
   \nonumber \\
   && \hspace*{2.9cm}
   \Bigl[ dn(n_x)K_x +dn(n_y)K_y +dn(n_z)K_z \Bigr] 
  /(\Lambda \Delta^{2}), 
\end{eqnarray}
\begin{eqnarray}
  && \hspace*{-0.3cm}
   \vec{K}(n_x,n_y,n_z) = \sum_{n_{1},n_{2},n_{3}} 
   (n_1,n_2,n_3) \Bigl\{ \exp \Bigl (-\Bigl( \pi/(\alpha L)  \Bigr)^{2} \Bigr) /\Lambda 
   \Bigr\} \times \nonumber \\
   && \hspace*{1.1cm} 
   \Bigl( sinc \Bigl( \frac{n_x+M_x n_1}{M_x} \Bigr) \Bigr)^{2P} 
   \Bigl( sinc \Bigl( \frac{n_y+M_y n_2}{M_y} \Bigr) \Bigr)^{2P} 
   \Bigl( sinc \Bigl( \frac{n_z+M_z n_3}{M_z} \Bigr) \Bigr)^{2P}   
\end{eqnarray}
\begin{eqnarray}
   && \hspace*{-1.3cm}
   \Delta(n_x,n_y,n_z) = \sum_{n_{1},n_{2},n_{3}} 
   \Bigl( sinc \Bigl( \frac{n_x +M_x n_1}{M_x} \Bigr) \Bigr)^{2P} \times
   \nonumber \\
   && \hspace*{3.1cm}
   \Bigl( sinc \Bigl( \frac{n_y +M_y n_2}{M_y} \Bigr) \Bigr)^{2P} 
   \Bigl( sinc \Bigl( \frac{n_z +M_z n_3}{M_z} \Bigr) \Bigr)^{2P}, 
\end{eqnarray}
\begin{eqnarray}
  && \Lambda= dn(n_{x})^2 +dn(n_{y})^2 +dn(n_{z})^2.  
\label{Longrg2}
\end{eqnarray}

The first Brillouin zone should take the summation of $-1 \leq n_{1} \leq 1$ (the degree is $P=3$), and M$_x$ is the mesh in the $x$ direction; the same procedures should be taken in the other directions due to the tetragonal crystal symmetry.
The sinc function is used to account for the long slopes of the $\vec{K}(n_x,n_y,n_z)$ and $\Delta(n_x,n_y,n_z)$ functions.
The index ranges for the $G(n_x,n_y,n_z)$ function are $0 \le n_x \le M_x/2$, $0 \le n_y \le M_y-1$ and $0 \le n_z \le M_z-1$, where M$_x$, M$_y$, and M$_z$ are the number of points in the $x, y$, and $z$ directions, respectively.





\end{document}